\documentclass{article} 
\usepackage[final]{colm2025_conference}

\usepackage{microtype}
\usepackage{hyperref}
\usepackage{url}
\usepackage{booktabs}

\usepackage{lineno}
\usepackage{graphicx}
\usepackage{amsmath}
\usepackage{array}
\usepackage[font=small,labelfont=bf]{caption}
\usepackage{titlesec} 

\definecolor{darkblue}{rgb}{0, 0, 0.5}
\hypersetup{colorlinks=true, citecolor=darkblue, linkcolor=darkblue, urlcolor=darkblue}

\title{Not Quite Anything: \\
Overcoming SAM’s Limitations for 3D Medical Imaging}

\author{Keith Moore \\
Deptartment of Biomedical Data Science \\
Stanford University \\
Stanford, CA 94304, USA \\
\texttt{kem1@stanford.edu}
}

\begin{document}

\ifcolmsubmission
\linenumbers
\fi

\maketitle

\begin{abstract}
Foundation segmentation models (such as SAM and SAM-2) perform well on natural images but struggle with brain MRIs where structures like the caudate and thalamus lack sharp boundaries and have poor contrast. 

Rather than fine-tune these models (e.g., MedSAM), we propose a compositional alternative where we treat the foundation model’s output as an additional input channel (like an extra color channel) and pass it alongside the MRI to highlight regions of interest.

We generate SAM-2 segmentation prompts (e.g., a bounding box or positive/negative points) using a lightweight 3D U-Net that was previously trained on MRI segmentation. However, the U-Net might have been trained on a different dataset. As such it's guesses for prompts are often inaccurate but often in the right region. The edges of  the resulting foundation segmentation \textit{guesses} are then smoothed to allow better alignment with the MRI. We also test prompt-less segmentation using DINO attention maps within the same framework.

This “has-a” architecture avoids modifying foundation weights and adapts to domain shift without retraining the foundation model. It achieves 96\% volume accuracy on basal ganglia segmentation, which is sufficient for our study of longitudinal volume change. Our approach is faster, more label-efficient, and robust to out-of-distribution scans. We apply it to study inflammation-linked changes in sudden-onset pediatric OCD.
\end{abstract}

\section{Introduction}

\begin{figure}[ht]
     \centering
     \includegraphics[width=.47\textwidth]{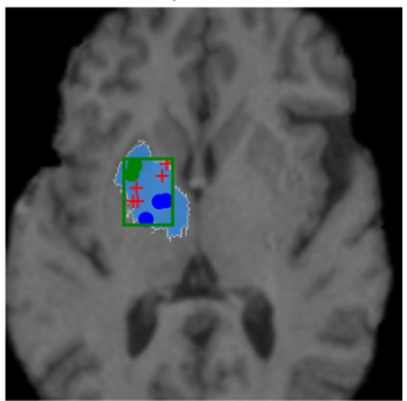} 
     \captionsetup{justification=centering}
     \caption{Poor isolation of caudate despite point prompts and bounding box 
     \newline (red-positive, blue-negative, and green-"don't care").}
     \label{fig:initial segmentation}
\end{figure}

Despite significant progress in segmenting natural images, foundation models like SAM \citep{37}, SAM-2 \citep{38}, and DINO \citep{20} fail to properly segment brain Magnetic Resonance Images (MRIs). The issue is that regions in the brain (particularly the subcortical regions) are difficult to distinguish due to limited contrast or clear visual boundaries. Retraining or fine-tuning the model on medical images helps (e.g., MedSAM \citep{38}) but the resulting segmentation is  inaccurate even with extensive prompting (see Figure~\ref{fig:initial segmentation} where we are trying to isolate the caudate with SAM).

We have had success using supervised models (such as 3D U-Net and UNETR) for brain segmentation \citep{moore2022}, but these models are sensitive to dataset distribution shift. We thought SAM might be doing poorly because it lacked 3D spatial information. In SAM-2, we tried encoding the Z dimension along the time dimension (i.e., treating the slices like frames of a movie), but this had no improvement to segmentation accuracy.  The other option of reachitecting SAM-2 to add the Z-axies and retrain on MRI images was impractical. This led us to explore a different strategy.

Instead of fine-tuning the foundation model, we use the existing inference engine to generate a 2D segmentation guess of each MRI slice as a second “channel” or color. This output is passed alongside the original MRI slice-by-slice into a multi-channel 3D U-Net to reconstruct the segmented volume. Our hypothesis was that the foundation model might help the U-Net find the intended structures.

In this architecture, the likely-inaccurate "guesses" by the foundation model act like an alternate imaging sequence (similar to T2 or FLAIR MRI imaging modalities), drawing attention to features in the T1 image. This “has-a” construction leverages the foundation model for landmark discovery, adapts to distribution shift, and remains fast to train with minimal supervision.

\subsection{Clinical Context and Motivation}

Our clinical objective is to analyze thousands of pediatric MRIs to investigate immune-related inflammation in neuropsychiatric disorders. We are particularly interested in the caudate and thalamus (small subcortical structures comprising approximately 20,000 voxels within an 8.6M voxel 3D image (240×240×155)). The effect we aim to measure is roughly a 10\% change in volume across two timepoints requiring a segmentation accuracy exceeding 94\% to distinguish signal from noise (see Appendix~\ref{appendix:threshold}).

Unlike segmentation tasks involving large lesions or tumors, our goal is to detect small volume changes in pre-defined structures. Even small boundary errors can obscure clinically meaningful differences, requiring models with higher edge precision than typical MRI segmentation tools. Real-world scans vary in quality, resolution, and orientation across sites. While improved instrumentation may help, clinical research must often work with heterogeneous and imperfect data. We sought methods that were robust to distribution shift, do not require retraining, and can operate at scale without extensive manual labeling.

\subsection{Prior Related Work}

Segmentation is often considered a \textbf{solved problem} in computer vision and rely on edge or texture cues to isolate structures. Brain MRI presents harder challenges where the relevant structures lack clear visual boundaries and exhibit low contrast with surrounding tissue (see Appendix~\ref{appendix:threshold}).

Finding the exact border is difficult even for a skilled radiologist.  In tasks such as tumor segmentation (e.g., BRATS21 \citep{29}), broader margins are often acceptable because precise boundaries are resolved histologically. Our use case, however, requires high precision in edges as a 10\% error is on average only 12 voxels per slice (i.e., 20,000/155).

\textbf{Traditional neuroimaging pipelines} such as ASEG \citep{25} and SAMSEG \citep{24} generate smooth, topologically consistent segmentations using atlas-based or probabilistic models, but they are computationally expensive (often requiring over an hour per scan). 

\textbf{CNN/Deep learning:} FastSurfer (\citep{henschel2020fastsurfer}) improves runtime performance using a 2.5D CNN-based architecture, but remains single-modal and can be sensitive to imaging artifacts and image distribution shift. Similarly, 3D U-Net \citep{ronneberger2015u} and its transformer cousin, UNETR \citep{27}, improve accuracy through improved spatial information but have a similar issue in sensitivity to  distribution shifts despite data augmentation\citep{moore2023}.

\textbf{Foundation models} (such as SAM \citep{37}, SAM-2 \citep{38} and DINO \citep{20}) have poor performance on medical images (see section~\ref{tab:baseline_results}. Others have fine-tuned these models on 2D medical images (e.g., MedSAM\citep{28} and  SAM-UNet\citep{27}); however, these have worked best in situations where there is clear boundaries between regions (i.e., MRIs of the abdomen). 

\textbf{Our approach:} These limitations led us to reframe the problem. Rather than treat segmentation as an “is-a” specialization of a foundation model (requiring full retraining), we adopt a compositional “has-a” strategy. We treat the foundation model as a frozen module that emits noisy guesses (used as an additional input channel) allowing a lightweight 3D U-Net to learn robust, high-accuracy segmentations from weak supervision.

\subsection{From Fine-Tuning to Composition}

We hypothesized that even flawed foundation model outputs could aid segmentation if we treated the segmentation "guess" as an additional input channel. This mirrors radiology practice, where multiple views (T1, T2, FLAIR) are integrated as channels to help resolve ambiguity. We treat a coarse segmentation from SAM or DINO as a synthetic modality and pair it with the raw MRI in a 3D multi-channel U-Net student model. The student learns to integrate this weak signal into an accurate 3D segmentation.

Functionally, the segmentation guess acts like a crude attention map. Architecturally, we shift from an \textit{is-a} inheritance model (where the foundation model is the solution) to a \textit{has-a} design (where the segmentation model \textit{has-a} foundation model as one of its components).

\section{System Architecture}

\begin{figure}[ht]
     \centering
     \includegraphics[width=1\textwidth]{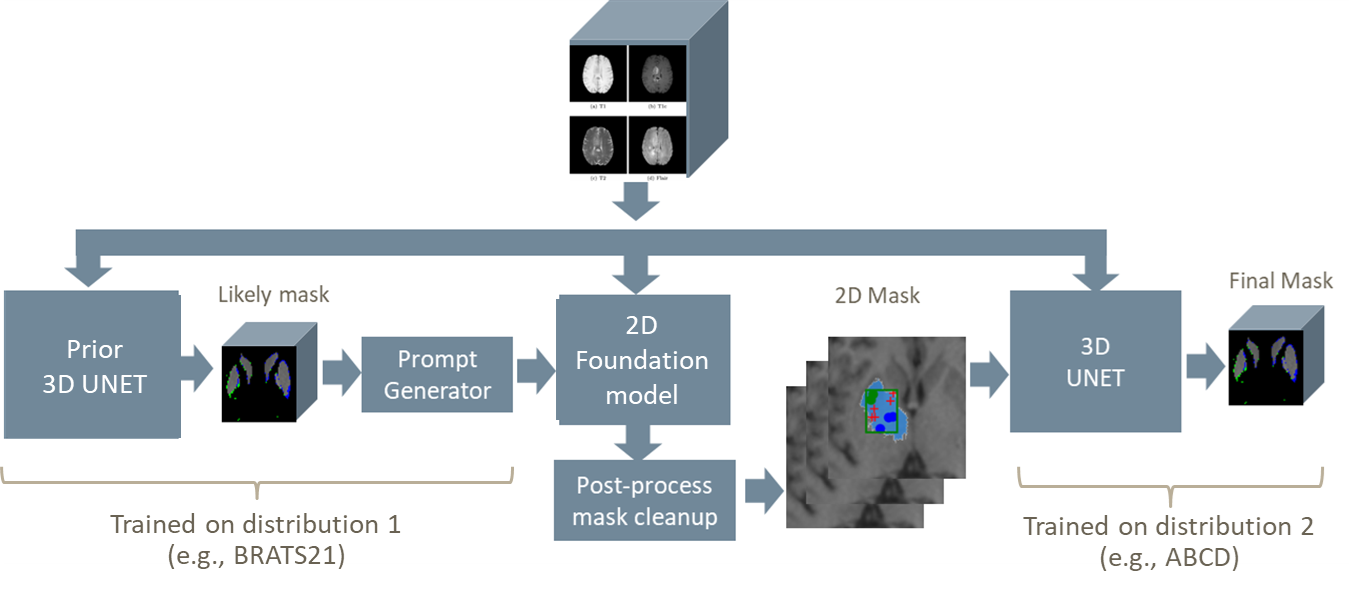} 
     \caption{System architecture. A prior U-Net prompts a 2D foundation model. Its mask guess is merged with a raw MRI and passed to a lightweight 3D U-Net student.}
     \label{fig:rough-architecture}
\end{figure}

Figure~\ref{fig:rough-architecture} outlines the system pipeline. At its core is a \textbf{foundation model} (SAM, SAM-2, MedSAM, or DINO) that produces 2D segmentation guesses. For prompt-based models, bounding boxes and point prompts are generated by a prior U-Net iteration (called the \textbf{prompt generator}). These prompts need not be precise. Rough bounding boxes are sufficient to trigger meaningful guesses.

The output mask is then post-processed and merged with the raw T1 MRI as a second channel, forming a 3D volume passed into a \textbf{3D U-Net student model}. This model refines the segmentation using both raw structure and foundation cues. The student model is small (~3M parameters) and fast to train (under 30 minutes for 100 MRIs). Earlier U-Nets can be re-used as prompt engines to bootstrap later training rounds.

\subsection{Segmentation Guess Construction}
Our pipeline produces a segmentation “guess” that serves as a second input channel to the 3D U-Net. This guess is derived either via prompted foundation models (e.g., SAM, SAM-2, MedSAM) or prompt-less attention models (e.g., DINO).

\begin{figure}[ht]
\centering
\includegraphics[width=.9\textwidth]{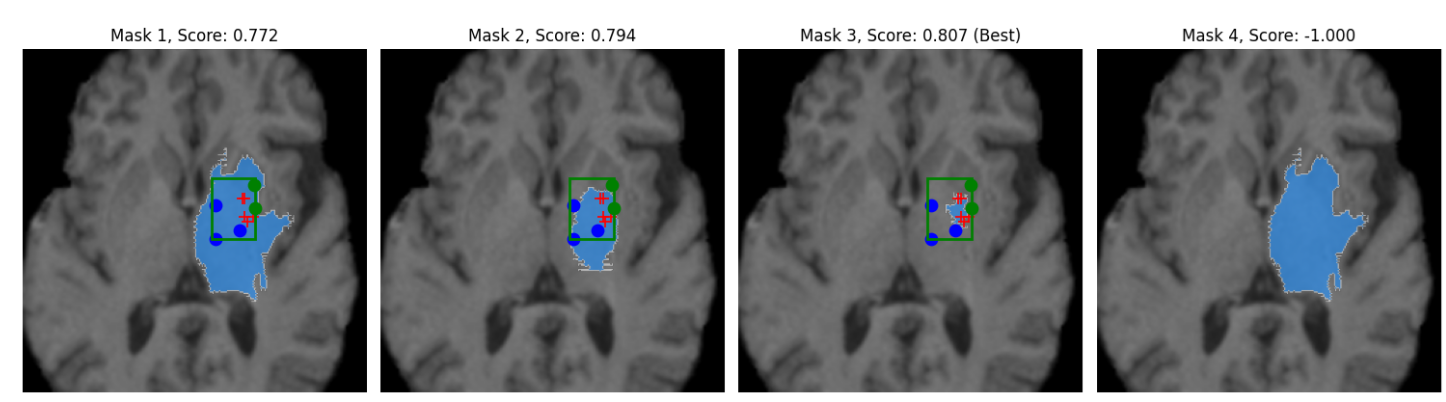}
\caption{Multiple SAM-2 segmentations for a given prompt}
\label{fig:sam points}
\end{figure}

\textbf{Prompt-based generation:} Using a prior U-Net iteration, we extracted bounding boxes and five labeled points per slice (foreground, background, or “don’t care”) to serve as prompts. Boxes were jittered for robustness. These were passed into the respective API for SAM, SAM-2, or MedSAM.\footnote{MedSAM from \url{medsam-vit-base}; SAM-2 from \url{sam2.1_hiera_large.pt}; DINO from \url{dino_vits8}.} Figure~\ref{fig:sam points} shows multiple segmentation masks returned by SAM-2 for one prompt.

\begin{figure}[ht]
\centering
\includegraphics[width=.9\textwidth]{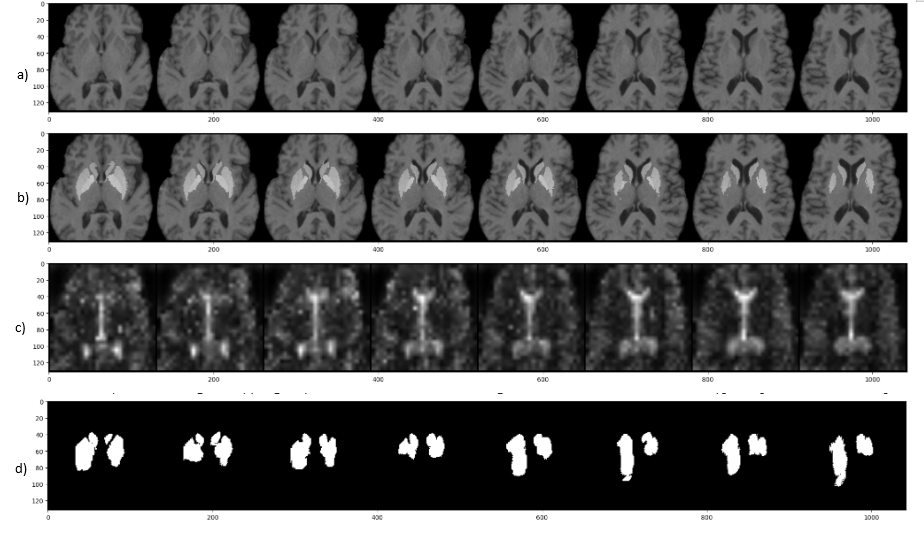}
\caption{a) T1 image, b) ground truth, c) DINO attention, d) SAM-2 mask}
\label{fig:dino-sam}
\end{figure}

\textbf{Prompt-less generation:} DINO attention maps were used directly as segmentation guesses, bypassing the prompt engine. These maps often resembled T2-weighted MRIs (Figure~\ref{fig:dino-sam}), supporting our “multi-modal” framing of foundation outputs as synthetic MRI contrasts.

\textbf{Post-processing:} For SAM-based models, we selected the highest-confidence mask per slice. Inspired by MedSAM, we tested contour-based filtering and found that selecting the largest contour within the bounding box was generally effective. To smooth boundaries, we applied a signed distance map using a clipped sigmoid blur (10-pixel range). This improved volume alignment despite minor DICE degradation.

\textbf{Volume reconstruction and augmentation: }Segmentation guesses were reassembled into 3D NifTi volumes stored alongside the raw MRIs. At training time, the T1 image and the segmentation guess were loaded as separate channels. Dataloaders supported optional substitution of the second channel (e.g., blank or T2 for baseline comparisons). Each subject volume was augmented 4x using MONAI transforms (RandomGamma, GaussianNoise, affine perturbations). The 3D U-Net trained on these dual-channel inputs.
\subsection{Datasets}
We used BRATS21 \citep{33} for initial training and validation. It includes 1,940 multi-modal MRIs (T1, T2, T1ce, FLAIR) from 140 patients, totaling 168,000 slices. Ground truth labels were generated in prior work using SAMSEG \citep{25}, though accuracy may degrade in tumor-distorted regions. A subset of ABCD \citep{34} was used to test distribution shift. 

\begin{table}[ht]
\small
\centering
\begin{tabular}{|>{\raggedright\arraybackslash}p{3.7cm}|>{\raggedright\arraybackslash}p{4.7cm}|>{\raggedleft\arraybackslash}p{1cm}|>{\raggedleft\arraybackslash}p{1cm}|>{\raggedleft\arraybackslash}p{1cm}|}
\hline
\textbf{Dataset} & \textbf{Description} & \textbf{Subjects} & \textbf{MRIs} & \textbf{\% Labels} \\
\hline
BRATS21 \citep{33}   & Tumor surgery  & 140 & 1,940 & 100\% \\
ABCD \citep{34}      & Pediatric longitudinal cohort & 11,857 & 45,678 & 20\% \\
\hline
\end{tabular}
\vspace{5pt}
\caption{Datasets used in this study; results focus on BRATS21 (Phase 1).}
\label{tab:datasets}
\end{table}

\section{Experiments Overview}

We organized our experiments into four stages to isolate the effects of architectural dimensionality, generalization, foundation model input, and signed distance smoothing:

\begin{enumerate}
  \item \textbf{2D/3D U-Net Baselines (Supervised):} 
  Includes a naive 2D U-Net (T1+T2), a 2D SAM run using U-Net-derived prompts (no refinement), and a full 3D U-Net (T1+T2) to test the benefit of volumetric context.

  \item \textbf{3D UNETR Baselines (Supervised):} 
  UNETR trained on BRATS21 (in-distribution) and then applied to ABCD scans (out-of-distribution) without fine-tuning to test brittleness under distribution shift.

  \item \textbf{3D U-Net + Foundation Guess (No Distance Map):} 
  Dual-channel input using T1 plus segmentation guesses from MedSAM, SAM-2, or DINO; used to assess weak priors without labels or retraining.

  \item \textbf{3D U-Net + Guess + Signed Distance Map:} 
  Same input as above, but the guess is softened at the edge  by using a signed distance map\citep{35} and a sigmoid falloff to smooth gradients and improve alignment despite noise.
\end{enumerate}
All U-Net variants used the same architecture and training pipeline, enabling clean comparison of structure, prior quality, and refinement.

\section{Results}
To establish labels, we used generated labels from SAMSEG and Freesurfer (two well established tools).  As we wanted to compare T1+T2 vs T1+SAM-2 guesses, we used SAMSEG as a common labeling tool.
\subsection{Baseline Performance}

Table~\ref{tab:baseline_results} reports performance from models without foundation model input. These baselines underscore the value of 3D context and expose the limits of both 2D segmentation and unrefined foundation models:

\begin{itemize}
  \item \textbf{2D U-Net (T1+T2)} generalized poorly from slices to volumes despite high training DICE, underscoring the limitations of 2D architectures.
  \item \textbf{2D SAM (Prompted)} outperformed the 2D U-Net on test DICE ($\sim$0.46), though this likely reflects prompt quality more than SAM's segmentation skill.
  \item \textbf{3D UNETR} trained on BRATS21 performed well in-distribution (DICE 0.76, Vol Acc 0.93) but degraded substantially on ABCD scans, highlighting brittleness under domain shift.
\end{itemize}

\begin{table}[ht]
\centering
\small
\begin{tabular}{|>{\raggedright\arraybackslash}p{3cm}|>{\raggedright\arraybackslash}p{5cm}|>{\raggedright\arraybackslash}p{1.5cm}|>{\raggedright\arraybackslash}p{1.cm}|>{\raggedright\arraybackslash}p{1cm}|}
\hline
\textbf{Model} & \textbf{Description} & \textbf{Validation DICE} & \textbf{Test DICE} & \textbf{Vol Acc}\\
\hline
\multicolumn{5}{|c|}{\rule{0pt}{2.5ex}\textbf{2D baseline (Supervised)}} \\ \hline
2D U-Net (T1 + T2) & Slice-wise supervised baseline & .72 & .31 & .83 \\ \hline
2D SAM (Prompted) & SAM mask compared to ground truth & N/A & .46 & .48 \\ \hline
\multicolumn{5}{|c|}{\rule{0pt}{2.5ex}\textbf{3D Baseline (Supervised)}} \\ \hline
3D UNETR (BRATS21) & Used for prompt generation; trained on BRATS21 (in distribution) & .92 & .76 & .93 \\ \hline
3D UNETR (ABCD) & Tested on ABCD data without fine-tuning (out-of-distribution) & .92 & .69 & .82 \\ \hline
\end{tabular}
\caption{Baseline performance of 2D and 3D models without foundation model inputs}
\label{tab:baseline_results}
\end{table}

\textit{Note:} The strong BRATS21 results were expected, but the distribution shift to ABCD proved substantial. This is likely due to anatomical differences between adult and pediatric brains.\footnote{In retrospect, BRATS tumors distort anatomy, but pediatric scans from ABCD showed unexpected structural variation.}

\subsection{Performance with Foundation Model Guesses}

Table~\ref{tab:initial_results} shows results using segmentation guesses from MedSAM, SAM-2, and DINO, each merged as a second channel with the T1 image. We evaluate both raw guesses and versions processed through a Signed Distance Map (SDM).

\begin{itemize}
  \item \textbf{Without SDM:} Results varied. SAM-2 improved test DICE to 0.75; DINO was less effective.
  \item \textbf{With SDM:} Volume accuracy improved across all models. SAM-2 + SDM achieved the best overall performance (DICE 0.76, Vol Acc 0.96).
\end{itemize}

\begin{table}[ht]
\centering
\small
\begin{tabular}{|>{\raggedright\arraybackslash}p{3.5cm}|>{\raggedright\arraybackslash}p{5cm}|>{\raggedright\arraybackslash}p{1.5cm}|>{\raggedright\arraybackslash}p{.75cm}|>{\raggedright\arraybackslash}p{.75cm}|}
\hline
\textbf{Model} & \textbf{Description} & \textbf{Validation DICE} & \textbf{Test DICE} & \textbf{Vol Acc}\\
\hline
\multicolumn{5}{|c|}{\rule{0pt}{2.5ex}\textbf{3D U-Net Without Signed Distance Map}} \\ \hline
3D U-Net (T1 only) & No auxiliary input & .44 & .46 & .40 \\ \hline
3D U-Net (T1 + T2) & T1 and T2 as channels & .64 & .66 & .77 \\ \hline
3D U-Net (T1 + MedSAM) & MedSAM as second channel & .70 & .69 & .80  \\ \hline
3D U-Net (T1 + SAM-2) & SAM-2 as second channel & .75 & .75 & .89  \\ \hline
3D U-Net (T1 + DINO) & DINO attention map as 2nd channel & .57 & .56 & .51 \\ \hline
\multicolumn{5}{|c|}{\rule{0pt}{2.5ex}\textbf{3D U-Net With Signed Distance Map}} \\ \hline
3D U-Net (T1 + T2 + SDM) & T2 smoothed with SDM & .58 & .62 & .65 \\ \hline
3D U-Net (T1 + MedSAM + SDM) & MedSAM guess smoothed with SDM & .71 & .71 & .91 \\ \hline
\textbf{3D U-Net (T1 + SAM-2 + SDM)} & \textbf{Best model (guess + SDM)} & \textbf{.76} & \textbf{.76} & \textbf{.96} \\ \hline
3D U-Net (T1 + DINO + SDM) & DINO guess smoothed with SDM & .61 & .62 & .65 \\ \hline
\end{tabular}
\vspace{1em}
\caption{Segmentation performance using foundation model guesses (with/without SDM)}
\label{tab:initial_results}
\end{table}

All models were trained with identical hyperparameters, random seeds, and 4x augmentation. 

\section{Discussion}

The central result is that segmentation guesses from a prompted SAM-2 model, when merged with the T1 image, significantly improve performance (especially when smoothed via a signed distance map (SDM)).\footnote{We had some experience with signed distance maps\citep{35} in a prior project \citep{moore2023} but were surprised at how effective it was for aligning edges between channels.}

\subsection{Impact of the Signed Distance Map}

The form of the SDM was critical. Initial tests used a broad soft gradient across the entire mask, which degraded performance (likely due to misalignment with DICE scores). Restricting the blur to a 10-voxel band around the boundary yielded substantially better results.

We adjusted our loss function as a dynamic loss (i.e., it started by prioritizing a DICE loss, then once the regions mostly overlap, the loss shifts to prioritizing the boundary loss).\footnote{Conceptually this is a $loss=\sigma(loss_1-.5)*(loss_1)+(1-\sigma(loss_1-.5))*loss_2$.}  Even with imperfect alignment, SDM provides a consistent optimization signal, leading to improved volume estimation.

\subsection{Edge Uncertainty Remains}

\begin{figure}[ht]
     \centering
     \includegraphics[width=1\textwidth]{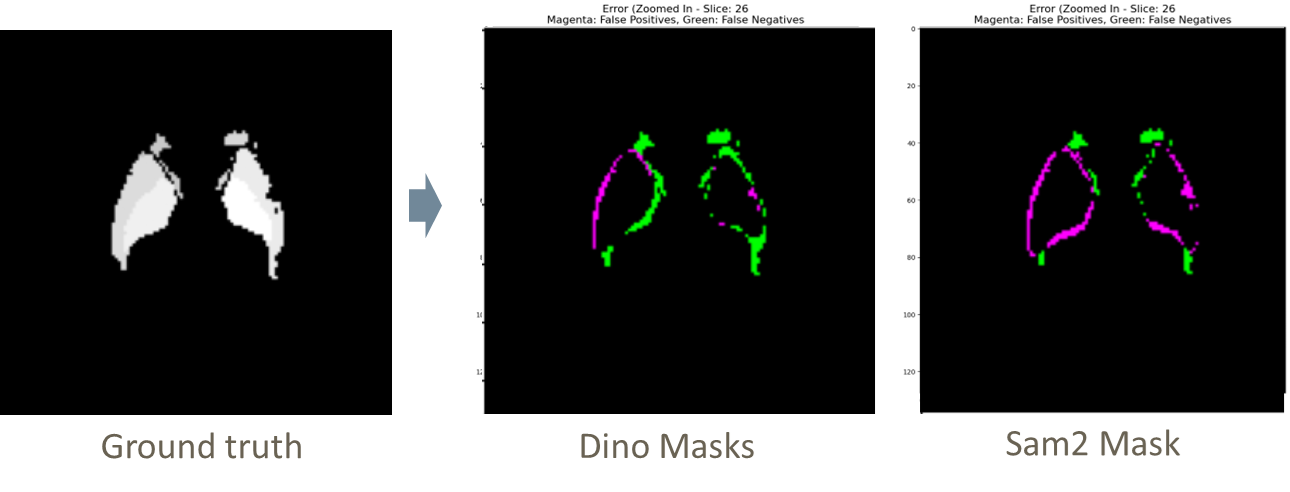} 
     \caption{Edge discrepancies: magenta = false positives, green = false negatives}
     \label{fig:edges}
\end{figure}

While SDM improves volume accuracy, precise edge alignment remains imperfect (Figure~\ref{fig:edges}). Most errors for SAM-2 and DINO occur within 1–2 pixels of the boundary.  This is not surprising given the low contrast of regions like the caudate.

Improving edge precision would likely require higher weighting on edge losses or significantly more training. Still, this may not be necessary: volume is cubic in radius, while surface area is quadratic. Small edge errors tend to have limited impact on volume estimates, which are our primary focus.

\subsection{Being Close is Good Enough}

A surprising result was how tolerant the student model was to noisy segmentation guesses. As long as the foundation model's output overlapped the general region, the 3D U-Net learned the correct structure. The guess appears to act as a weak “attention” signal, guiding spatial focus even though the architecture lacks explicit attention.

\begin{figure}[ht]
     \centering
     \includegraphics[width=.9\textwidth]{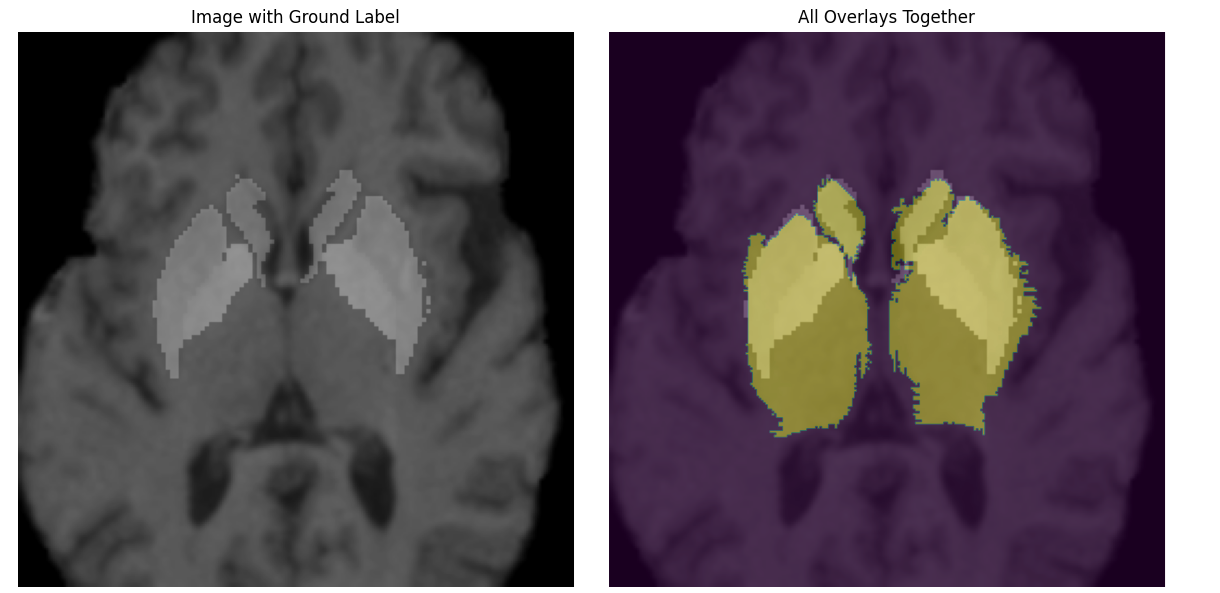} 
     \caption{SAM-2 guess overlaps with target region but is not precise.}
     \label{fig:overlay}
\end{figure}

Figure~\ref{fig:overlay} shows an example where the SAM-2 mask overlaps the target region but includes irrelevant areas. Despite this, the U-Net correctly localizes the structure, likely by learning spatial correlations across slices.

\subsection{Fine-Tuning vs Adaptation}

The architecture resembles a teacher-student setup: the teacher (e.g., UNETR + SAM-2) provides noisy inference-only supervision, and the student (3D U-Net) is trained on these outputs. No gradients are passed to the teacher, avoiding the cost of fine-tuning.

This adaptation strategy lets us wrap foundation model outputs with domain-specific processing (e.g., SDM) without unlocking layers or training new weights. It is faster, simpler, and more flexible.

Training was efficient, and the student generalized well (even with noisy guesses) by learning distribution shift. A small number of labeled samples served as anchors, while the second channel guided spatial attention.

This hybrid setup, combining frozen foundation models with structural smoothing and a lightweight student, is promising for low-supervision segmentation tasks.

\section{Conclusion and Next Steps}

This project showed that even inaccurate guesses from foundation models can support accurate 3D segmentation when properly adapted. Fine-tuning MedSAM added little; adapting outputs worked better (perhaps due to recovering spatial information). Prompt generation from a prior student model was effective, and DINO worked as a prompt-free option (although not as well). Using a clipped signed distance map consistently improved volume accuracy, especially for weaker inputs.

The key insight is that \textbf{adapting foundation model outputs} (rather than modifying foundation model weights) can deliver robust segmentation. Whether this generalizes remains open, and next steps includes ablation studies, but early results on pediatric ABCD data are promising.

Our goal was to enable volumetric tracking of basal ganglia in adolescents to study inflammation-linked OCD phenotypes. The next step is to apply this model to real clinical scans and assess its ability to detect volume shifts linked to inflammation.
\section{Acknowledgements}
The author would like to recognize the guidance of Prof. Serena Yeung-Levy, Xiaohan Wang and James Burgess of the MARVL lab at Stanford for their help and guidance on the project. 
\newpage
\bibliography{references1}
\bibliographystyle{colm2025_conference}

\appendix
\renewcommand{\thesection}{\Alph{section}}
\titleformat{\section}
  {\normalfont\Large\bfseries}
  {Appendix \thesection:}{1em}{}

\renewcommand{\thesubsection}{\thesection.\arabic{subsection}}

\newpage
\section {Supplemental Material}
\subsection{Issues in intensity}
The figure below shows the complexity of determining the label based on intensity. The intensities overlap for nearby regions.  Brain Atlases tend to do better here as they use the probabilities that a pixel is part of one region or another by looking at the adjacent pixels.  Contrast dyes are used in MRIs to try to help separate regions; however, the contrast dyes have some risk and even when used these regions are difficult to discern.  The reality is that ground truth is exceptionally difficult with high variance by radiologists on where the exact edge is located in an MRI (making volume measurements quite difficult).
\begin{figure}[ht]
  \centering
  \includegraphics[width=.9\textwidth]{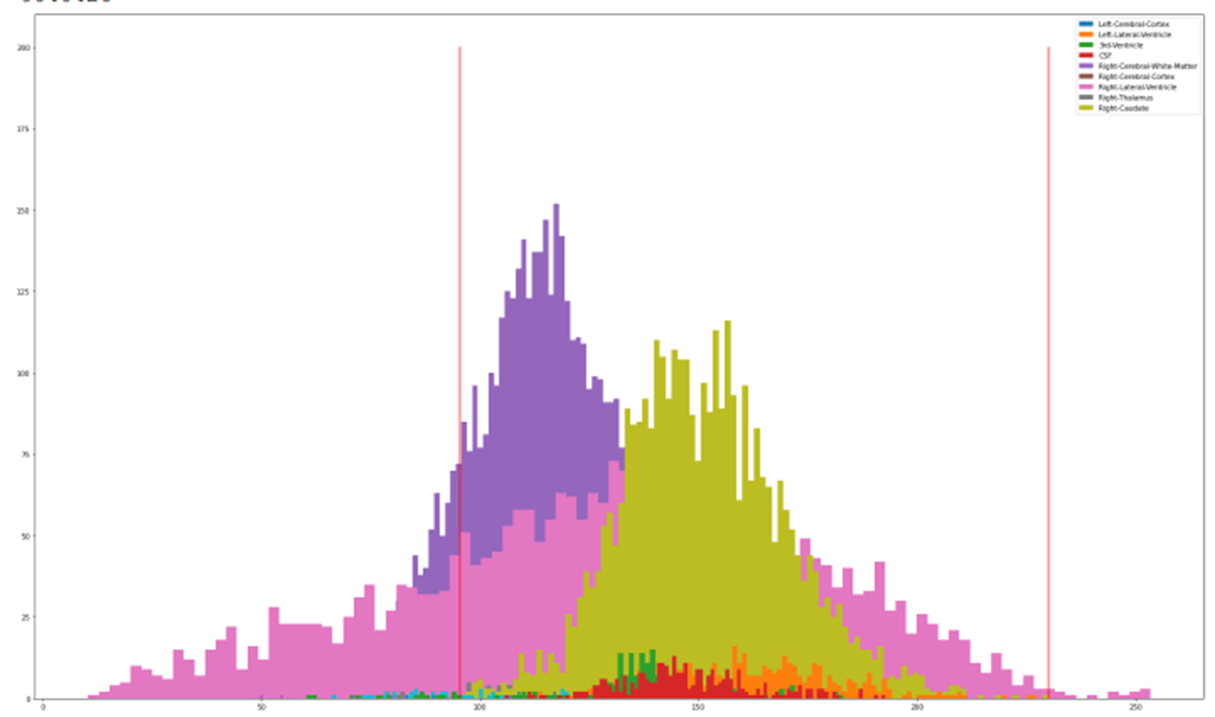} 
  \caption{MRI pixel intensities overlap for brain regions making segmentation difficult}
  \label{fig:overlap}
 \end{figure}

 \subsection{DICE score improved through selective classification}
 There are outliers in the BRATS21 dataset (as can be seen in supplementary figure 8).  Here we note that three of the volumes have very strong differences in labels from the "ground truth".  Analyzing the data, we see that in one of the MRIs a tumor has strongly distorted the region of interest making it unrecognizable as being the basal ganglia.  The other issues were due to some blurring in the image acquisition.  This is common and can, to some degree, be corrected.

 \begin{figure}[ht]
  \centering
\includegraphics[width=.5\textwidth]{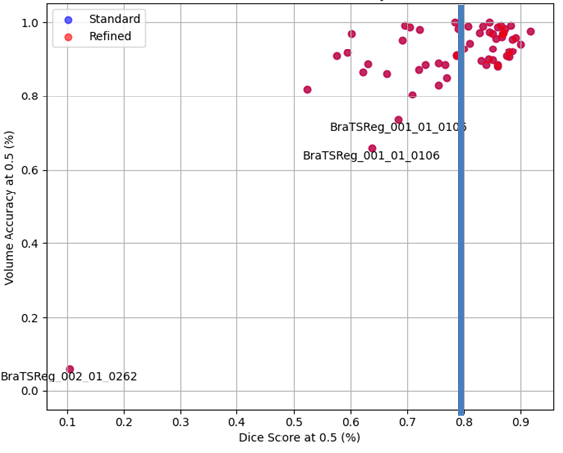} 
  \begin{minipage}{0.9\textwidth}
    \caption{
     Outliers in the DICE distribution for BRATS21
    }
  \end{minipage}
  \label{fig:saliency}
\end{figure}
\newpage
 \subsection{Comparison of 2D, 3D U-Net and UNETR volume accuracy}
 Overall the performance of the UNETR did best with a 93\% volume accuracy when fine-tuned on the BRATS21 distribution. The 3D U-Net also had good scores having only 3M parameters. Note, the study below is when using 4 radiology modes that have been aligned (T1, T2, T1ce and FLAIR)

\subsection{DINO does pretty well}
Supplemental Figure 10 shows a comparison of the DINO initial guess and the U-Net refirement.  The final row shows the error to the target labels.  This is frankly an amazing result (i.e., it is very, very close) and it is possible that this result would improve with a stronger focus in the loss function or a lower learning rate.  Next step would be to conduct those experiments.  This figure is moved here as it wasn't the main focus of the paper.

\begin{figure}[ht]
  \centering
  \includegraphics[width=0.9\textwidth]{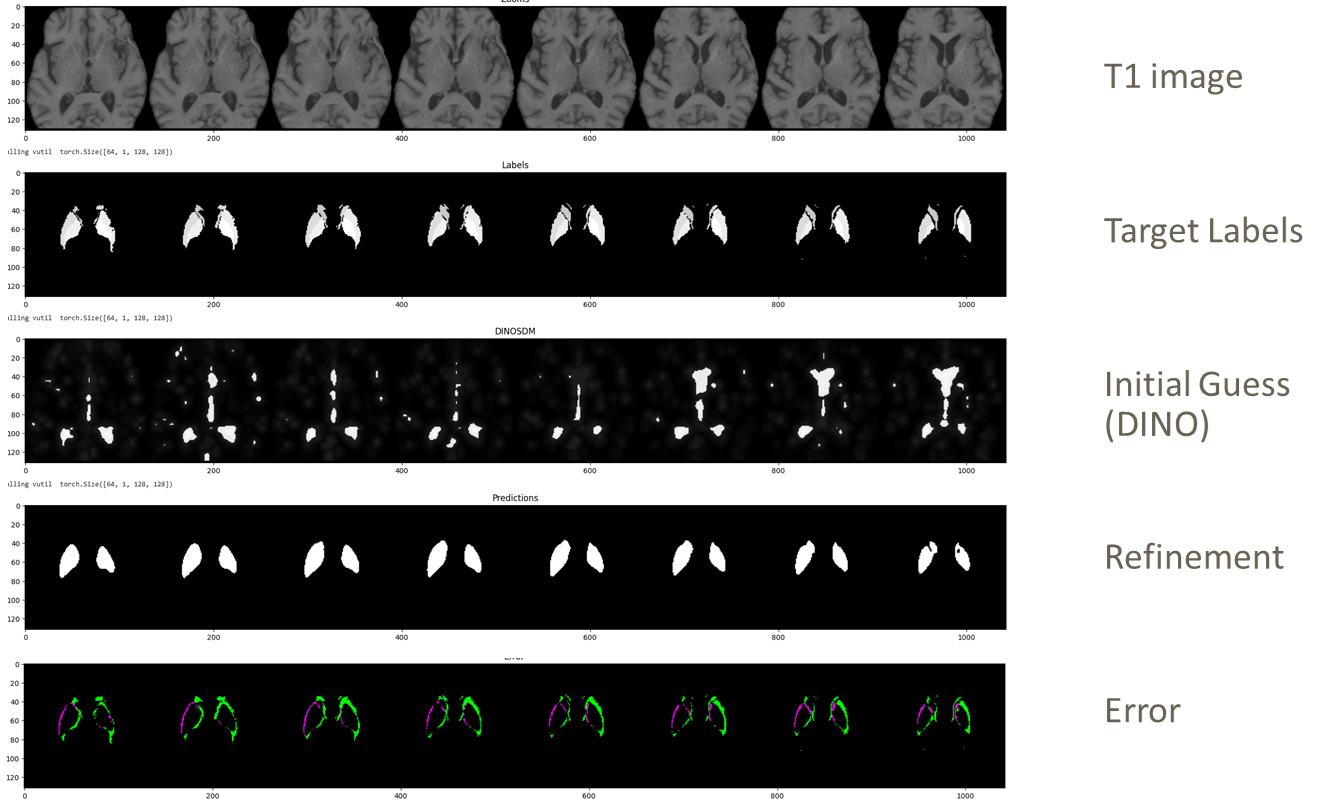} 
  \caption{Result of refinement after DINO's initial guess.}
  \label{fig:segmentation}
\end{figure}

\subsection{Volume Measurement Accuracy for Detecting Effect Size}
\label{appendix:threshold}

To determine the required volume measurement accuracy necessary to detect a 10\% effect size in volume changes between cases and controls, we analyze the relationship between measurement error, sample size, and the detectable effect size.

\paragraph{Key Variables}
\begin{itemize}
    \item $V_{\text{avg}}$: Average volume of the brain region (e.g., caudate).
    \item $\Delta V_{\text{cases}}$: Average per-patient volume change in the \textbf{case} group.
    \item $\Delta V_{\text{controls}}$: Average per-patient volume change in the \textbf{control} group.
    \item $\Delta V_{\text{diff}} = \Delta V_{\text{cases}} - \Delta V_{\text{controls}}$: True effect size (10\% of $V_{\text{avg}}$).
    \item $\epsilon_\Delta$: Measurement error in detecting volume change over two time points.
    \item $\text{SE}_{\text{diff}}$: Standard error of the difference between the groups.
    \item $n_{\text{cases}}, n_{\text{controls}}$: Sample sizes for cases and controls, respectively.
    \item $Z$: Critical $Z$-score for the desired confidence level (e.g., 1.96 for 95\% confidence, 1.645 for 90\% confidence).
\end{itemize}

\paragraph{Variance from Measurement Error:}
The variance of the volume change per patient due to measurement error is additive, as the volume is measured at two time points:
\[
\text{Variance from measurement error per patient} = 2 \cdot (\epsilon_\Delta \cdot V_{\text{avg}})^2.
\]
Thus, the total measurement error variance for each group is:
\[
\sigma_{\text{meas}}^2 = \frac{2 \cdot (\epsilon_\Delta \cdot V_{\text{avg}})^2}{n_{\text{cases}}} + \frac{2 \cdot (\epsilon_\Delta \cdot V_{\text{avg}})^2}{n_{\text{controls}}}.
\]

\paragraph{Standard Error of the Difference:}

\[
\text{SE}_{\text{diff}} = \sqrt{\sigma_{\text{meas}}^2}.
\]

\paragraph{Condition to Detect Effect Size}
The true difference must satisfy:
\[
\Delta V_{\text{diff}} \geq Z \cdot \text{SE}_{\text{diff}}.
\]

Substitute $\text{SE}_{\text{diff}}$:
\[
\Delta V_{\text{diff}} \geq Z \cdot \sqrt{\frac{2 \cdot (\epsilon_\Delta \cdot V_{\text{avg}})^2}{n_{\text{cases}}} + \frac{2 \cdot (\epsilon_\Delta \cdot V_{\text{avg}})^2}{n_{\text{controls}}}}.
\]

Rearranging for $\epsilon_\Delta$ gives:
\[
\epsilon_\Delta \leq \frac{\Delta V_{\text{diff}}}{Z \cdot V_{\text{avg}} \cdot \sqrt{\frac{2}{n_{\text{cases}}} + \frac{2}{n_{\text{controls}}}}}.
\]

Substitute $\Delta V_{\text{diff}} = 0.10 \cdot V_{\text{avg}}$:
\[
\epsilon_\Delta \leq \frac{0.10}{Z \cdot \sqrt{\frac{2}{n_{\text{cases}}} + \frac{2}{n_{\text{controls}}}}}.
\]

\paragraph{Example Calculation}
Assume:
\begin{itemize}
    \item $n_{\text{cases}} = 240$,
    \item $n_{\text{controls}} = 10,000$.
\end{itemize}

First, calculate:
\[
\frac{2}{n_{\text{cases}}} + \frac{2}{n_{\text{controls}}} = \frac{2}{240} + \frac{2}{10,000}.
\]

\[
\frac{2}{n_{\text{cases}}} = 0.00833, \quad \frac{2}{n_{\text{controls}}} = 0.0002.
\]

Add these:
\[
\frac{2}{n_{\text{cases}}} + \frac{2}{n_{\text{controls}}} = 0.00853.
\]

Take the square root:
\[
\sqrt{0.00853} = 0.0923.
\]

\paragraph{At 95\% Confidence (Z = 1.96)}
Substitute $Z = 1.96$:
\[
\epsilon_\Delta \leq \frac{0.10}{1.96 \cdot 0.0923}.
\]

\[
\epsilon_\Delta \leq \frac{0.10}{0.1809} \approx 0.553.
\]

Thus, the required measurement accuracy is:
\[
1 - \epsilon_\Delta \geq 1 - 0.0553 = 94.47\%.
\]

\paragraph{At 90\% Confidence (Z = 1.645)}
Substitute $Z = 1.645$:
\[
\epsilon_\Delta \leq \frac{0.10}{1.645 \cdot 0.0923}.
\]

\[
\epsilon_\Delta \leq \frac{0.10}{0.1517} \approx 0.659.
\]

Thus, the required measurement accuracy is:
\[
1 - \epsilon_\Delta \geq 1 - 0.0659 = 93.41\%.
\]

\paragraph{Conclusion}
To reliably detect a 10\% effect size in volume changes:
\begin{itemize}
    \item At 95\% confidence, the required measurement accuracy is approximately \textbf{94.47\%}.
    \item At 90\% confidence, the required measurement accuracy is approximately \textbf{93.41\%}.
\end{itemize}

\end{document}